\documentclass[sigconf, nonacm]{acmart}
\usepackage{subcaption}
\usepackage{multirow}

\usepackage{arydshln}

\usepackage{pifont}

\AtBeginDocument{%
  }

\begin{document}

\title{A Switch-Centric In-Network Architecture for Accelerating LLM Inference in Shared-Memory Network}

\author{Aojie Jiang}
\email{aojie@smail.nju.edu.cn}
\affiliation{%
  \institution{Nanjing University}
  \country{China}
}

\author{Kang Zhu}
\email{kang_zhu@smail.nju.edu.cn}
\affiliation{%
  \institution{Nanjing University}
  \country{China}
}

\author{Zhiheng Zhang}
\email{zhz@smail.nju.edu.cn}
\affiliation{%
  \institution{Nanjing University}
  \country{China}
}

\author{Zhengxu Su}
\email{suzhengxu@smail.nju.edu.cn}
\affiliation{%
  \institution{Nanjing University}
  \country{China}
}

\author{Juntao Liu}
\email{liujuntaoyjy@chinamobile.com}
\affiliation{%
  \institution{China Mobile Research Institute}
  \country{China}
}

\author{Yuan Du}
\email{yuandu@nju.edu.cn}
\affiliation{%
  \institution{Nanjing University}
  \country{China}
}

\author{Li Du}
\email{ldu@nju.edu.cn}

\affiliation{%
  \institution{Nanjing University}
  \country{China}
}

\begin{abstract}
Tensor parallelism (TP) has become a key technique for latency-sensitive LLM inference, but it introduces frequent, tightly synchronized All-Reduce operations that lie directly on the inference critical path, making inter-GPU communication a major performance bottleneck.
In-network computing offers a promising approach to mitigating this bottleneck.
However, we observe that the existing solution, NVLink SHARP (NVLS), is fundamentally constrained by its GPU-driven execution model that exposes in-network computing as GPU-issued, element-granular memory operations rather than a switch-orchestrated collective.
This design forces redundant switch-GPU-switch data movement and cannot perform scale-aware quantized All-Reduce due to its lack of cross-element coordination.

To address these limitations, we propose SCIN, the first switch-centric in-network computing architecture for multi-accelerator shared-memory networks.
SCIN replaces GPU-driven collective orchestration with an in-switch accelerator (ISA) that directly initiates memory transactions and orchestrates the complete All-Reduce operation. 
This switch-driven execution eliminates redundant data movement inherent in NVLS, shortens synchronization paths, and fully offloads collective execution from GPUs.
Furthermore, centralized ISA control enables cross-element coordination between quantized data and their scale factors. SCIN exploits this capability through a pipelined dequantization-reduction-requantization datapath, enabling scale-aware in-network quantized All-Reduce and nearly halving communication volume with negligible accuracy degradation.
We implement SCIN on a multi-FPGA prototype to validate its feasibility. 
For a simulated 8-GPU system, SCIN accelerates small- and large-message All-Reduce by up to 1.8x and 2.6x over NVLS, respectively, translating into up to 1.12x TPOT and 1.42x TTFT speedups.
\end{abstract}


\maketitle

\section{Introduction}

The rapid growth of large language models (LLMs) \cite{jiang2024mixtralexperts,llama2} has imposed unprecedented computation and memory demands, making distributed inference \cite{pope2023efficiently,agrawal2023sarathi} increasingly essential in modern multi-accelerator systems \cite{nvl72,ubmesh,lpunetwork}.
Among existing parallelization strategies, tensor parallelism (TP) \cite{narayanan2021efficient,zhang2026towards} shards individual layers across multiple GPUs and is increasingly used not only to accommodate models that exceed the memory capacity of a single GPU, but also to reduce inference latency by parallelizing layer computations in latency-sensitive serving scenarios \cite{wang2026moebiusservingmixtureofexpertmodels,nvidia2026moedensegemm}, such as in agentic systems \cite{10849561}.
TP can therefore be valuable even when the model fits on a single GPU.
However, TP introduces frequent, tightly synchronized All-Reduce operations that lie directly on the critical path and are often difficult to overlap with computation in low-latency inference scenarios where there is insufficient compute to hide communication \cite{gond2026tokenweaveefficientcomputecommunicationoverlap, taneja2026sifarsynchronizationfreeallreducelowlatency}.
As GPU compute throughput continues to outpace interconnect bandwidth, inter-GPU collective communication is becoming an increasingly important bottleneck in TP inference.

NVIDIA DGX nodes \cite{ishii2018nvswitch}, powered by high-bandwidth NVLink and NVSwitch \cite{danskin2016pascal,ishii2022nvlink}, have become key infrastructure for scaling up LLM systems via TP.
To further reduce collective communication overhead, NVIDIA introduced NVLink SHARP (NVLS) \cite{nvidia_dgx_docs,klenk2020network}, which performs reduction and multicast operations inside NVSwitches.
NVLS improves upon software-based Ring All-Reduce by offloading reduction and multicast to NVSwitch. Nevertheless, TP All-Reduce remains a substantial bottleneck with NVLS, accounting for up to 49\% of prefill time and 19\% of decode time in our profiling on an H200 system under low-concurrency configurations.
Motivated by this overhead, we examine NVLS's collective execution path and identify a fundamental architectural limitation: 
\textit{NVLS employs a GPU-driven collective execution model in which the switch passively processes GPU-issued, element-granular memory operations rather than orchestrating the collective as a whole}.
This execution model has two direct consequences: it forces redundant switch--GPU--switch data movement and precludes scale-aware quantized All-Reduce.

\textbf{First, NVLS's GPU-issued load/store semantics inherently cause redundant switch--GPU--switch data movement during All-Reduce.}
NVLS initiates in-network reduction through GPU-issued load instructions. To preserve load semantics, the switch-produced result must first return to the requesting GPU instead of being broadcast directly. The GPU must then issue store instructions to send the result back to the switch for broadcast. This inherent switch--GPU--switch detour increases latency and wastes bandwidth.

\textbf{Second, NVLS's element-granular reduction model precludes scale-aware quantized All-Reduce.}
For example, with symmetric quantization, each FP16/BF16 value is represented as $r_i=q_i s_i$, where $q_i$ is a quantized low-precision value, such as FP8, and $s_i$ is its scale factor. The correct reduction is therefore $\sum_i q_i s_i$, whereas NVLS can compute only $\sum_i q_i$, because its element-granular reduction model lacks the cross-element coordination required to associate quantized data with their corresponding scale factors.
Section \ref{limitations} explains these limitations in detail.
This limitation forces NVLS All-Reduce to operate at BF16/FP16 in mainstream LLM serving systems \cite{kwon2023efficient,zheng2024sglang,tensorrtllm}, even though prior work \cite{lee2024exploring,xiao2023smoothquant} has shown that 8-bit activation quantization is often sufficient to preserve model accuracy. This precision mismatch prevents NVLS from realizing the bandwidth savings offered by model quantization and highlights a significant opportunity to reduce communication overhead through scale-aware low-precision in-network computing.

These limitations motivate a shift from NVLS's GPU-driven collective execution model to a switch-driven one. We propose SCIN, a switch-centric in-network computing architecture for multi-accelerator shared-memory networks. 
\textit{SCIN's key insight is to move collective orchestration from GPU-issued load/store instructions to an in-switch accelerator (ISA), which directly orchestrates All-Reduce execution.}

This switch-driven execution addresses both limitations of NVLS. First, because the ISA itself initiates the memory transactions for All-Reduce, it can broadcast reduced results directly to participating GPUs as soon as reduction completes, eliminating NVLS's redundant switch--GPU--switch path and fully offloading collective orchestration from GPU SMs. During collective execution, GPUs only synchronize with the ISA to signal input readiness and await output visibility. Because each GPU synchronizes directly with the ISA over a single network hop, SCIN reduces barrier overhead relative to NVLS, which is especially beneficial for small-message All-Reduce.
Second, unlike NVLS, the switch-centric paradigm allows the ISA to centrally coordinate data and scale factor fetching, buffering, and computing.
Accordingly, we augment the ISA with a pipelined dequantization--reduction--requantization datapath and maintain the association between quantized operands and their scale factors, thereby enabling scale-aware quantized All-Reduce. This in-network quantization (INQ) nearly halves the communication volume with negligible accuracy degradation.
We co-design the switch microarchitecture, protocol extensions, and system-level execution support for switch-driven collectives, and implement SCIN on a multi-FPGA prototype to validate its feasibility.

To the best of our knowledge, SCIN is both the first work to propose a switch-centric in-network architecture for multi-accelerator shared-memory systems and the first to leverage in-network computing to accelerate quantized All-Reduce in LLM inference. Our main contributions are as follows:

\begin{itemize}
\item 
We identify a fundamental architectural limitation of NVLS: its GPU-driven execution model exposes in-network reduction as GPU-driven, element-granular memory operations, inherently causing redundant switch--GPU--switch data movement and precluding scale-aware quantized All-Reduce.

\item We propose SCIN, the first switch-centric in-network computing architecture for multi-accelerator shared-memory networks, 
where an in-switch accelerator (ISA) orchestrates All-Reduce to eliminate redundant data movement, reduce synchronization overhead, and enable scale-aware quantized All-Reduce.
\item We implement SCIN on a multi-FPGA prototype and evaluate it with a hardware-calibrated simulation of an eight-H200 system, achieving up to 1.8$\times$ and 2.6$\times$ speedups over NVLS for small- and large-message All-Reduce, respectively, and translating into up to 1.12$\times$ TPOT and 1.42$\times$ TTFT speedups.
\end{itemize}

\begin{figure}
    \centering
    \begin{subfigure}{\linewidth}
        \centering
        \includegraphics[width=0.9\linewidth]{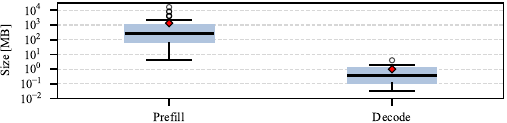}
        \caption{Distribution of All-Reduce message sizes across varying input configurations in LLaMA-2-70B TP inference (Prefill vs. Decode; batch size $\in \lbrack 1, 128\rbrack$, sequence length $\in \lbrack 128, 4096\rbrack$).}
        \label{fig2:a}
    \end{subfigure}

    \begin{subfigure}{\linewidth}
        \centering
        \includegraphics[width=0.9\linewidth]{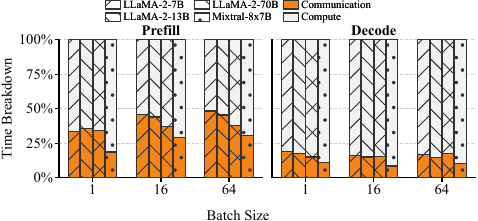}
        \caption{Computation and communication time breakdown of LLaMA-2 and Mixtral inference on eight NVIDIA H200 GPUs using TensorRT-LLM at FP8 inference precision with TP=8 and NVLS enabled.}
        \label{fig2:b}
    \end{subfigure}
\caption{Communication characteristics of tensor-parallel LLM inference.}    \label{fig:2}
\end{figure}

\section{Background and Motivation}
\begin{figure*}[!t]
  \centering
  \includegraphics[width=\textwidth]{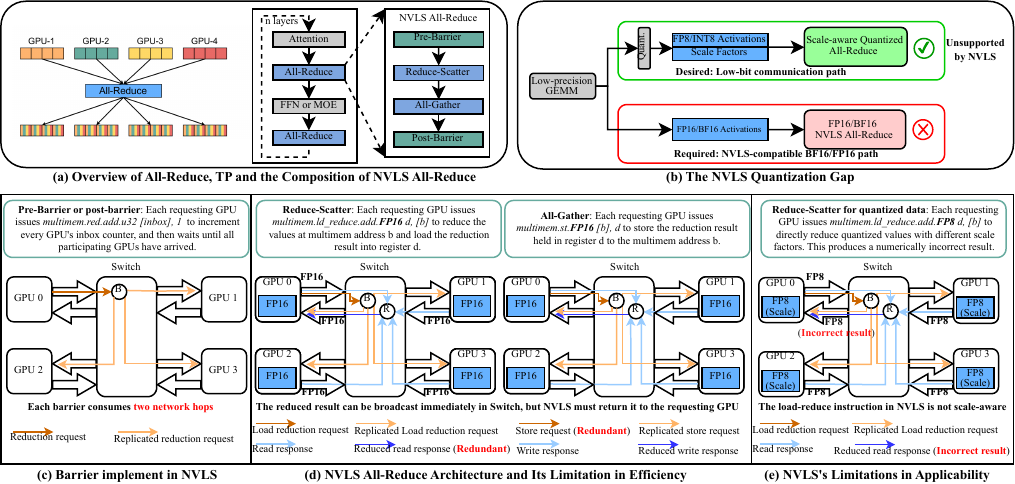}
  \caption{Motivation for SCIN: TP All-Reduce, the NVLS quantization gap, and NVLS architectural limitations (Only GPU 0's memory transfers are shown).}
  \label{fig:fig1}
\end{figure*}

\subsection{Tensor Parallelism in LLM Inference}
\label{sec:tp}

\textit{1) Communication Characteristics of TP Inference:} Under TP, each transformer layer typically performs two All-Reduce operations: one after the attention block and one after the MLP block. At FP16/BF16 precision, the size of each All-Reduce is $2bsh$ bytes during prefill but only $2bh$ bytes during each decode step, where $b$, $s$, and $h$ denote the batch size, prefill sequence length, and hidden dimension, respectively. Thus, for the same $b$ and $h$, a prefill collective carries $s$ times more data than a decode collective, as illustrated in Fig.~\ref{fig2:a}. Meanwhile, autoregressive decoding repeatedly invokes these small collectives at every layer for each generated token. Therefore, the prefill and decode stages impose distinct interconnect requirements \cite{ma2026challenges}. Prefill involves bandwidth-intensive, large-message All-Reduce operations, whereas decode invokes numerous latency-sensitive, small-message All-Reduce operations at every token-generation step. Consequently, the interconnect must provide both high bandwidth and low latency to optimize TTFT and TPOT, respectively.

\textit{2) The Communication Bottlenecks in TP Inference:} All-Reduce in TP inference lies directly on the critical path, forcing each accelerator to stall until the collective operation completes. Prior approaches to mitigating All-Reduce overhead, such as kernel fusion \cite{wu2025miragemultilevelsuperoptimizertensor} and microbatching \cite{zhu2025nanoflowoptimallargelanguage}, rely on overlapping communication with computation from large batches. However, such large batch sizes are typical only in training or high-concurrency inference scenarios, but not in low-latency inference serving, where there is insufficient compute to hide communication \cite{gond2026tokenweaveefficientcomputecommunicationoverlap, taneja2026sifarsynchronizationfreeallreducelowlatency}. To the best of our knowledge, none of the open-source serving systems such as vLLM \cite{kwon2023efficient}, SGLang \cite{zheng2024sglang} and TensorRT-LLM \cite{tensorrtllm} enable compute-communication overlap by default for TP inference. To quantify this limitation, we profile the communication overhead on a real H200 system using the TensorRT-LLM v1.3.0. As shown in Fig.~\ref{fig2:b}, under FP8 inference with TP=8, NVLS-accelerated All-Reduce still accounts for up to 49\% and 19\% of execution time in the prefill and decode stages, respectively. 
We expect this communication bottleneck to become increasingly prominent as recent advances in software systems continue to eliminate many non-communication bottlenecks and compute throughput increasingly outpaces interconnect bandwidth. For example, with Megakernel optimization~\cite{spector2025look}, SiFAR~\cite{taneja2026sifarsynchronizationfreeallreducelowlatency} reports that NVLS All-Reduce accounts for approximately 30\% of decoding time. Moreover, software optimizations improved GPT-J~\cite{gpt-j} performance on H100 by $2.4\times$--$2.9\times$ over the six months between MLPerf 3.1 and 4.0~\cite{reddi2019mlperf,softspeedup}.
On the hardware side, for example, from the NVIDIA A100 \cite{nvidiaA100} to the B200 \cite{nvidiaB200}, peak dense FP16 performance increased from 312 TFLOPS to 2.25 PFLOPS, a $7.2\times$ increase, whereas NVLink bandwidth grew from 600 GB/s to 1.8 TB/s, only a $3\times$ increase.

\textit{3) Communication Compression for TP Inference:} To mitigate the communication bottleneck in TP inference, many approaches have explored quantizing the data communicated during All-Reduce. One class of prior work \cite{low-bit,hansen2024communication,li2024flash} is based on direct All-Reduce \cite{9238637}, where each GPU directly sends its quantized activations to all other GPUs and then performs local reduction after dequantization. However, direct All-Reduce scales poorly with GPU count \cite{tacos}. 
Another representative design, EQuARX \cite{ahmed2025equarx}, quantizes the data communicated in Ring All-Reduce over Google TPU torus networks \cite{jouppi2023tpuv4opticallyreconfigurable}. 
When quantization is applied to Ring All-Reduce with N GPUs, 
the data in each chunk must undergo $N-1$ rounds of quantization during its reduce-scatter stage,
causing quantization errors to accumulate in each round. As a result, although EQuARX adopts a dynamic block-wise quantization strategy, it still suffers from substantial accuracy degradation due to the fundamental limitation of the Ring algorithm.

\subsection{Limitations in Existing In-Network Computing Architecture}
\label{limitations}

The high cost of All-Reduce has motivated the development of specialized hardware to accelerate collective communication.
The existing solution, NVIDIA's NVLink SHARP (NVLS) \cite{klenk2020network,nvidia_dgx_docs}, integrates in-network computing into NVSwitches. As shown in Fig.~\ref{fig:fig1}, NVLS All-Reduce has four phases: (1) a pre-barrier ensures that all input data is ready; (2) during Reduce-Scatter, each GPU pulls the operands for its assigned partition, while the switch reduces responses associated with each request and returns the result to the requester; (3) during All-Gather, each GPU broadcasts its reduced partition to all peers; and (4) a post-barrier ensures that all writes are globally visible. By offloading reduction and broadcast, NVLS reduces data movement and communication overhead compared with software Ring All-Reduce \cite{klenk2020network}.
\textit{However, we observe that NVLS is fundamentally limited by its GPU-driven execution model that exposes in-network computing as GPU-issued, element-granular memory operations, with the switch passively processing these operations.}

\textbf{First, NVLS's GPU-issued load/store semantics inherently cause redundant switch--GPU--switch data movement during All-Reduce.}
As shown in Fig.~\ref{fig:fig1}(d), in-network reduction in NVLS is initiated by a GPU-issued load-style instruction (\path{multimem.ld_reduce}). When the resulting reduction request reaches the switch, it is replicated and forwarded to all participating GPUs. Each target GPU memory system services the request as a regular load and returns a response to the switch, which reduces the response values associated with the same request. To preserve the semantics of the load instruction, the reduced result in the switch must be returned to the requesting GPU and written to its destination register, rather than being immediately broadcast to all participating GPUs. To complete the subsequent All-Gather phase, the requesting GPU must issue additional store instructions (\path{multimem.st}) that send the result back to the switch for broadcast. Consequently, the reduced data traverses the switch-to-GPU and GPU-to-switch path before being distributed. This detour is inherent to NVLS's load-based reduction interface, introducing additional latency and wasting link bandwidth.

\textbf{Second, NVLS's element-granular reduction model precludes scale-aware quantized All-Reduce.}
Broadcast is agnostic to data representation: the switch can simply forward quantized values and their scale factors as opaque data. In-network reduction, however, must account for their numerical relationship. Under symmetric quantization, each FP16/BF16 value is represented as $r_i=q_i s_i$, where $q_i$ is a low-precision value, such as FP8 or INT8, and $s_i$ is its scale factor. Therefore, before reduction, each quantized value must be dequantized using its corresponding scale to a common higher-precision representation (FP16/BF16). The correct reduction is therefore $\sum_i q_i s_i$, whereas NVLS can compute only $\sum_i q_i$, which has no meaningful interpretation because the source operands have different scale factors, as shown in Fig.~\ref{fig:fig1}(e). Supporting scale-aware reduction in NVLS would require more than integrating quantization-related hardware; it would likely entail a fundamental architectural redesign.
NVLS exposes reduction as an address-based, element-granular memory operation, with elements at contiguous addresses within each packet reduced independently. Scale-aware in-network reduction, by contrast, requires coordination across elements at different addresses to associate each quantized value with its corresponding scale factor before dequantization, followed by accumulation and requantization. Such cross-element, metadata-coupled processing does not fit NVLS's existing per-element reduction abstraction.
Integrating it into NVLS would require a cross-layer redesign of the instruction interface, request and completion protocols, packet formats, and switch-side execution and state management. The system-level feasibility and cost-effectiveness of such a redesign remain unclear.

Consequently, even the latest Rubin-generation NVLS \cite{nvidia2026ptx94} still cannot perform scale-aware in-network reduction.
This limitation prevents NVLS from keeping pace with the evolving acceleration requirements of current inference systems. Specifically, current inference systems exhibit a growing precision mismatch between computation and All-Reduce communication, as illustrated in Fig.~\ref{fig:fig1}(b). Model quantization has been widely adopted to accelerate computation and reduce memory usage \cite{lang2024comprehensive,liu2023llm} and prior work \cite{lee2024exploring,xiao2023smoothquant} shows that 8-bit activation quantization can often preserve model accuracy. Therefore, quantized All-Reduce naturally aligns with these quantization pipelines. However, because NVLS lacks scale-aware reduction, frameworks such as vLLM, SGLang, and TensorRT-LLM must retain activations in FP16/BF16 for NVLS All-Reduce, preventing them from realizing the substantial bandwidth savings offered by model quantization.

Beyond these primary limitations, NVLS's GPU-driven reduction architecture introduces high synchronization overhead. Its system-level barriers synchronize GPUs through the switch, resulting in a two-hop synchronization path. In contrast, SCIN synchronizes each GPU directly with the centrally located ISA over only a single network hop. 
Moreover, NVLS does not fully offload collective communication to the switch: numerous GPU threads are still required to orchestrate massive payloads in parallel. Saturating the available link bandwidth typically requires 6 SMs \cite{bachan2025nccl227,gond2026tokenweaveefficientcomputecommunicationoverlap}, approximately 5\% of the SM resources on an H100.

\subsection{SCIN Design Philosophy}

\begin{figure*}[!t]
  \centering
  \includegraphics[width=\textwidth]{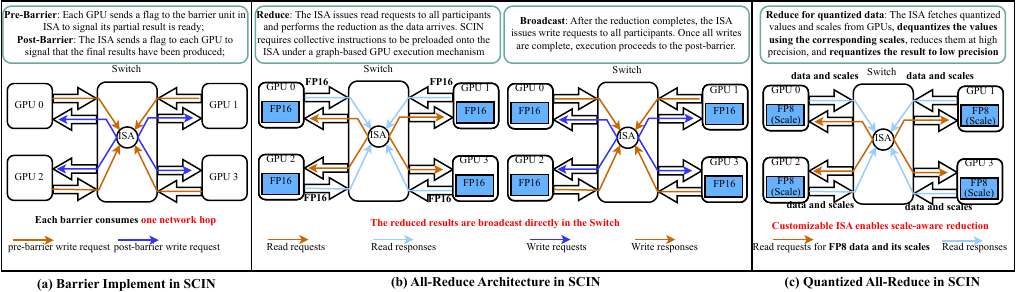}
  \caption{The architecture of SCIN.}
  \label{fig:fig3}
\end{figure*}



\begin{table}[t]
  \centering
  \caption{Comparison of NVLS and SCIN.}
  \label{tab:nvls-scin-comparison}

  \setlength{\tabcolsep}{3pt}
  \renewcommand{\arraystretch}{1.1}

  \begin{tabular}{@{}lccccc@{}}
    \toprule
    \multirow{2}{*}{Method} &
    \multirow{2}{*}{\shortstack{Bandwidth-efficiency\\ factor}} &
    \multicolumn{3}{c}{Network hops} &
    \multirow{2}{*}{\shortstack{Quantized\\All-Reduce}} \\
    \cmidrule(lr){3-5}
      & & Total & Data & Barriers & \\
    \midrule
    NVLS & $\frac{N}{N+1}$ & 12 & 8 & 4 & \ding{55} \\
    SCIN & $1$             &  6 & 4 & 2 & \ding{51} \\
    \bottomrule
  \end{tabular}
\end{table}

To overcome NVLS's limitations, we propose SCIN, a switch-centric in-network computing architecture for shared-memory AI-accelerator fabrics, as shown in Fig.~\ref{fig:fig3}. 
\textbf{SCIN's key insight is to move collective orchestration from GPU-issued load/store instructions to an in-switch accelerator (ISA), which directly orchestrates All-Reduce execution.} This switch-driven design eliminates NVLS's redundant data movement by exploiting the switch's central position and enables quantized All-Reduce by specializing the ISA for scale-aware reduction, unlocking the acceleration potential of in-switch computing.

To realize SCIN, we co-design the switch microarchitecture, protocol extensions, and system-level execution support for switch-driven collectives. 
An SCIN All-Reduce operation proceeds in four phases: pre-barrier, reduction, broadcast, and post-barrier. Through dedicated interfaces to the switch ports, the ISA can directly access the memory of attached accelerators for in-network data processing. Because the ISA, rather than the GPUs, initiates memory transactions for All-Reduce, it can broadcast reduced results back to participating accelerators immediately after reduction, eliminating redundant data movement in NVLS and enabling full offload of collective execution. 
During collective execution, GPUs only need to synchronize with the ISA to signal input readiness and await All-Reduce completion. 
Moreover, switch-centric execution allows the ISA to centrally coordinate data fetching, buffering, and in-network processing. Accordingly, we augment the ISA with a pipelined dequantization--reduction--requantization datapath and maintain the association between quantized operands and their source-specific scale factors, thereby enabling scale-aware quantized All-Reduce. Finally, moving collective initiation from GPUs to the ISA shortens each barrier synchronization path from GPU--switch--GPU, which requires two network hops, to GPU--ISA or ISA--GPU, which requires only one. This reduction in synchronization overhead is particularly beneficial for latency-sensitive, small-message All-Reduce.

Table~\ref{tab:nvls-scin-comparison} quantitatively compares SCIN with NVLS. For large messages, following the bandwidth-factor formulation used by NCCL \cite{nccltests}, we also use this bandwidth-efficiency factor $\eta$ to characterize how efficiently each architecture converts physical bus bandwidth into useful All-Reduce bandwidth:
\begin{equation}
B_{\mathrm{AllReduce}}
=
B_{\mathrm{bus}} \times \eta_{\mathrm{}}.
\end{equation}
In NVLS, its redundant round-trip data movement requires each GPU to transmit an additional reduced shard for which it serves as the requesting GPU. Assuming that an All-Reduce payload is evenly partitioned across $N$ GPUs, this shard accounts for $1/N$ of the payload, increasing the normalized traffic volume from $1$ to $1+1/N$. The resulting bandwidth-efficiency factor is therefore
\begin{equation}
\label{eq:nvls_bandwidth_efficiency}
\eta_{\mathrm{NVLS}}
=
\frac{1}{1+1/N}
=
\frac{N}{N+1}.
\end{equation}
SCIN eliminates such redundant transfer and thus achieves $\eta_{\mathrm{SCIN}}=1$. For small messages, we compare the total number of network hops on the critical path of a complete All-Reduce operation, including barrier synchronization.
SCIN halves both the data path and the synchronization path compared with NVLS, theoretically enabling up to a 2× speedup for small-message All-Reduce.

\section{Design and Implementation}
\label{sec:Design}
\begin{figure*}[!t]
  \centering
  \includegraphics[width=0.8\textwidth]{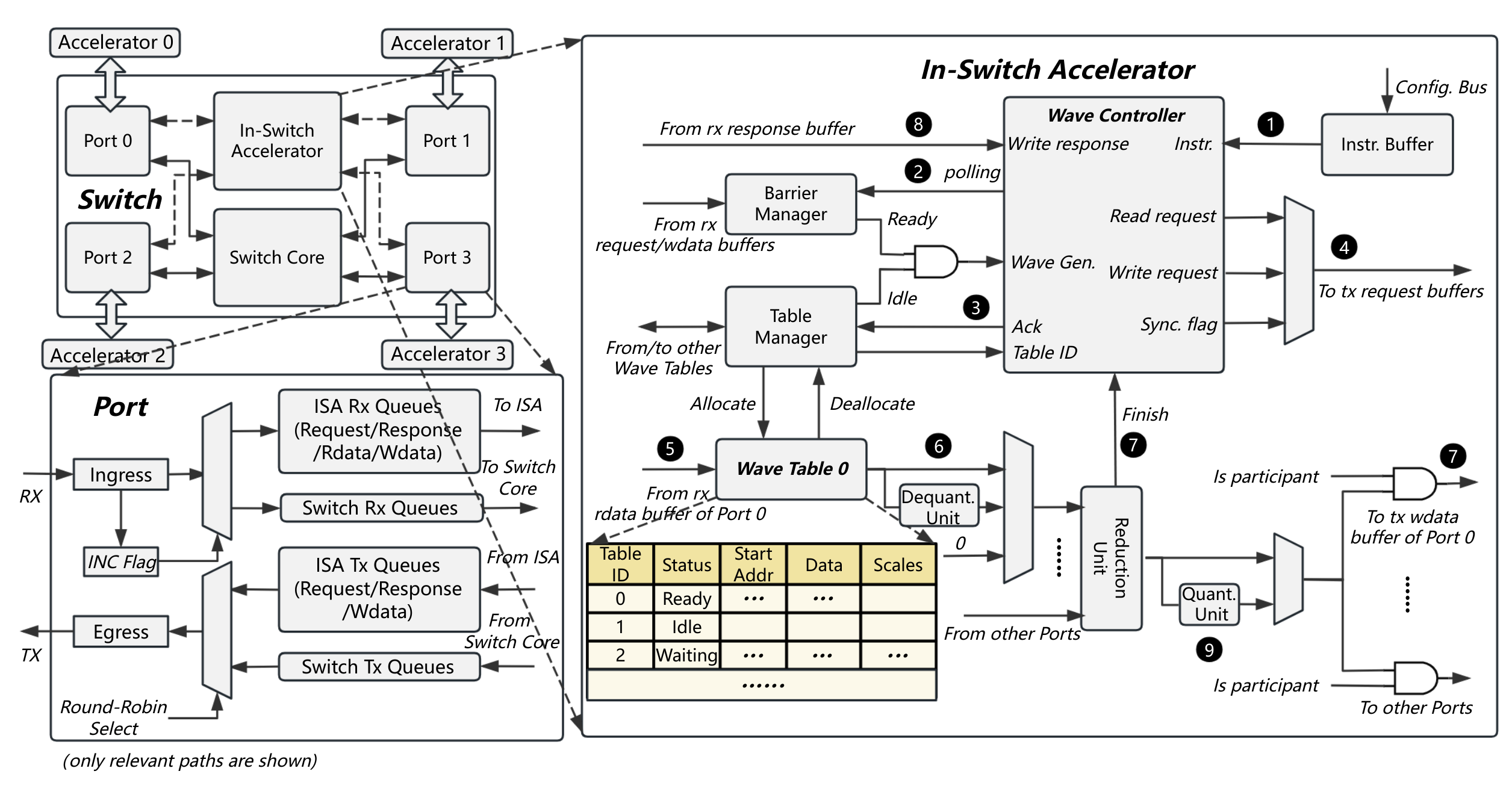}
  \caption{Switch Microarchitecture for SCIN.}
  \label{fig:4}
\end{figure*}

\subsection{Implementation Challenges}
1) Existing communication protocols and their hardware implementations are primarily designed to forward packets between endpoint accelerators, rather than allowing a switch-resident DMA-like engine to initiate memory transactions. The ISA should access attached accelerator memory without disrupting regular network functions and with minimal modifications to existing protocols and hardware.

2) Because All-Reduce lies directly on the critical path of TP inference, synchronization and runtime configuration delays directly increase inference latency. SCIN should provide lightweight synchronization and low-overhead ISA instruction configuration, while integrating with existing accelerator software through limited, localized modifications.

3) AI accelerators expose high degrees of concurrency, making packet arrival order at the switch nondeterministic. To sustain high bandwidth with limited on-chip resources, the ISA should efficiently manage concurrent out-of-order packet streams while isolating concurrent collectives.

In the following sections, we describe how SCIN addresses these challenges, analyze its scalability, and detail its microarchitecture and multi-FPGA prototype implementation.

\subsection{Network Protocol Extension}

To allow the ISA to access endpoint accelerator memory, we propose a hardware microarchitecture for the switch port, as illustrated in Fig.~\ref{fig:4}(left). In this architecture, the ISA is directly connected to each switch port to communicate with all accelerators within the network. 
We adopt a transport-layer organization used in prior designs such as UALink \cite{UALink} and OISA \cite{oisa}, with a dedicated queue for each message type. Each port maintains separate queues for memory-semantic messages, including Read/Write Data Queues, Request Queues, and Response Queues. We physically expand these queues into two independent sets: Switch RX/TX Queues and ISA RX/TX Queues. These queues independently serve standard packet forwarding and ISA-related packet forwarding, respectively. Because traffic through the ISA queues is explicitly scheduled by the ISA and its payloads are buffered in the ISA's internal tables, these ISA queues serve only as interfaces between the ISA and the transport layer, incurring negligible area overhead. The ISA communicates with the endpoint accelerators by issuing and receiving requests or responses strictly through its dedicated queues. We only introduce a 1-bit In-Network Computing (INC) flag into the packet header to support SCIN with negligible packet-header overhead. This flag is used to explicitly indicate whether a packet originates from or is destined for the ISA. On the receiving (RX) side, the port logic utilizes this 1-bit INC flag to demultiplex and route the incoming data streams to either the ISA or the Switch Core. On the transmitting (TX) side, data packets from both the ISA and the Switch Core are allocated transmission bandwidth using a round-robin arbitration scheme, thereby minimizing interference with regular network packets forwarding.

\subsection{Execution Mechanism}

SCIN's system-level execution support comprises two components: graph-based ISA instruction configuration and ISA--GPU synchronization. 1) Graph-based ISA instruction configuration: SCIN targets graph-based GPU execution models, such as NVIDIA CUDA Graphs \cite{nvidia_cuda_graphs} and AMD HIP Graphs \cite{amd_hip_graphs}, which are commonly used by modern LLM serving frameworks \cite{kwon2023efficient,zheng2024sglang}. 
A GPU graph captures GPU operations and their dependencies and packages them into a reusable executable that can be executed with a single graph launch, thereby reducing host-side kernel-dispatch overhead.
SCIN extends this graph-based execution model to ISA operations. During graph capture, the SCIN-extended runtime identifies All-Reduce operations in the computation graph, translates them into ISA instruction descriptors, and preloads the descriptors into the ISA instruction buffer. During execution, the GPU and the ISA therefore advance autonomously according to their predefined dependencies and synchronization events. For a given instantiated graph, the addresses, shapes, and sizes of intermediate activation tensors remain unchanged across replays. The ISA can then read GPU-produced operands directly from their producer buffers and write the reduced results to predetermined destination buffers, avoiding staging buffers and copies.
This design removes ISA configuration from the inference critical path, requires only localized runtime extensions, and exposes no explicit ISA-management interface to applications.

2) ISA--GPU synchronization: As shown in Fig.~\ref{fig:fig3}(a), SCIN uses pre-barrier and post-barrier phases for ISA--GPU synchronization. During the pre-barrier, each GPU makes its partial result globally visible, sets its ISA-resident arrival flag, and then polls a local completion flag. Once all arrival flags are set, the ISA initiates All-Reduce. During the post-barrier, the ISA waits for the operation and all result writes to complete, ensuring that the results are globally visible in GPU memory, and then updates the completion flags to release the GPUs. 

\subsection{In-switch Accelerator Design}

\label{sec:isa}
Building on the above substrate, we describe how the ISA performs in-network operations. Fig.~\ref{fig:4}(right) illustrates the ISA architecture.

\noindent\textbf{1. Wave-Based Regulation.} Because response packets may return in arbitrary order, the data arriving at the ISA may not be reduced immediately. Therefore, before issuing read requests, the ISA must reserve buffer space in advance for data returned in subsequent responses to avoid overflow. The size of the reduction buffer must be properly provisioned to sustain full link bandwidth, which can be estimated using Little's Law \cite{little2008little}.
For a single-tier topology, let the accelerator--switch bandwidth be $B$, the one-way link latency be $L$, and the accelerator response latency be $L_{\mathrm{acc}}$. The minimum reduction-buffer capacity required by SCIN is
\begin{equation}
\label{eq:scin_buffer}
C_{\mathrm{SCIN}} = B\left(2L + L_{\mathrm{acc}}\right).
\end{equation}
In comparison, the minimum buffer capacity required by an NVLS-like design is
\begin{equation}
\label{eq:nvls_buffer}
C_{\mathrm{NVLS}} = B\left(4L + L_{\mathrm{acc}}\right).
\end{equation}
Because the ISA uses a deterministic accumulation order and thus produces reproducible floating-point results, we provision the NVLS-like baseline with the minimum buffer capacity required to provide the same reproducibility guarantee, ensuring a fair comparison. By moving reduction initiation from GPUs to the ISA, SCIN reduces the round-trip latency of each reduction request from $4L + L_{\mathrm{acc}}$ to $2L + L_{\mathrm{acc}}$, thereby lowering the reduction-buffer capacity required to sustain full bandwidth. Section~\ref{sec:Hardware} evaluates the impact of buffer capacity on performance.

Given finite on-chip buffering, the ISA must regulate its injected traffic to prevent overflow while sustaining full bandwidth. Similar to NVLS~\cite{klenk2020network}, SCIN divides each request into multiple waves. Given a total buffer capacity $C$, the ISA allows up to $k$ waves to remain outstanding concurrently and reserves $C/k$ buffer capacity for each wave.  Synchronization is required between waves to ensure that the next wave is injected only after buffer resources used by the previous wave have been released. The ISA overlaps multiple waves in time, hiding synchronization latency while preserving buffer safety.

\begin{figure}
    \centering
    \includegraphics[width=\linewidth]{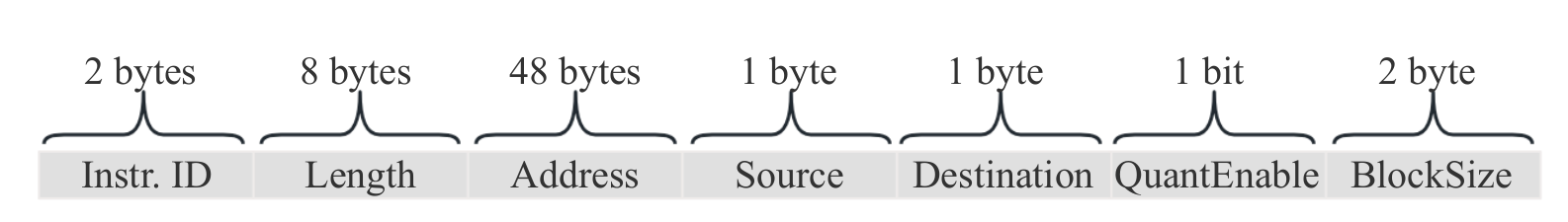}
    \caption{Instruction format of the ISA.}
    \label{fig:5}
\end{figure}

\noindent\textbf{2. Instruction Format.}
Fig.~\ref{fig:5} shows the instruction format for an 8-accelerator node. Each instruction contains an Instruction ID, a Length field specifying the transfer size, and a 48-bit source Address for each accelerator; results are written back in place. SCIN may optionally use NVLS's multimem mechanism to encode multiple accelerator buffers with a single address, shortening instructions at the cost of additional software memory management. The Source and Destination bitmasks identify the accelerators contributing operands and receiving results, respectively. Setting both masks to all ones specifies All-Reduce, whereas selecting all sources and one destination specifies Reduce. QuantEnable indicates quantized All-Reduce. Because activation data and scale factors are stored separately, they are loaded by two instructions; when QuantEnable is set, the next instruction in the buffer provides the per-accelerator scale-factor addresses.
BlockSize denotes the number of elements that share a common scale factor.

\noindent\textbf{3. ISA Microarchitecture.} The ISA consists of two key components: a wave controller and a set of per-accelerator wave tables. The wave controller is mainly responsible for fetching ISA instructions, performing synchronization, allocating storage, and issuing read/write requests. Each entry in the wave table stores one wave, including its data, base address, associated scale factor (when INQ is enabled), and state bits such as IDLE, WAITING, and READY.

Execution proceeds at wave granularity. After fetching an ISA instruction \ding{182}, the wave controller first polls the barrier manager to determine whether all participating accelerators have produced the required data \ding{183}. At the same time, it queries the table manager to determine whether any free entries are available in all relevant wave tables \ding{184}. Once both conditions are met, the controller issues read requests opportunistically based on the number of currently idle entries \ding{185}. If the available capacity is insufficient to cover the full transfer length, only a partial segment is issued, and the remaining segments are deferred until entries are released.
To ensure that each returning read response knows where it should be placed in the wave tables, the ISA reuses the existing tag field in the packet header to encode the location in the wave tables. Because the response packet carries the same tag as the corresponding request, the ISA can identify its storage position directly upon arrival. Concretely, the tag encodes the target table-entry index together with the packet offset within that entry. Since the tag (or transaction ID) field is typically large enough to distinguish all outstanding packets required to sustain full bandwidth, it is sufficient to uniquely identify all packets stored in the wave tables.
Upon arrival, each response packet is stored into the corresponding wave-table location according to its tag \ding{186}. Once all entries associated with a given wave become READY, the corresponding data is read out from the wave tables and fed to the reduction unit \ding{187}, after which the corresponding entries are released and made available for subsequent waves. Since the compute units in the ISA have a fixed processing latency, the wave controller can issue the corresponding write requests at the appropriate time along with the write data produced by the tree-based reduction unit \ding{188}, with the request addresses derived from the base addresses stored in the entries. Once all write responses are received \ding{189}, the operation is complete and the result becomes visible to the participating accelerators. The ISA then signals completion by writing a synchronization flag to the designated synchronization address of each participant \ding{185}. Notably, multiple synchronization resources can be integrated to support multiple parallel computation streams, and multiple wave controllers can also be incorporated to improve instruction-level parallelism.

\noindent\textbf{4. Implementation of Quantized All-Reduce.}
SCIN applies dynamic block-wise INT8 quantization to All-Reduce inputs and outputs. Each block contains BlockSize elements along the hidden dimension and uses an independent scale derived from its maximum magnitude. This achieves $1.94\times$ compression with negligible accuracy loss when BlockSize is 64 (Section~\ref{sec:INQ}). Unlike Ring-based schemes requiring $N-1$ quantization steps, INQ introduces only one quantization step during communication, independent of the GPU count $N$.

During execution, the wave controller first loads a fixed-size chunk of scale factors into the scale field of the allocated table entry---for example, 128 B when the wave size is 4 KB---and then loads the corresponding activation data. Once the wave entries become READY, they are read from the wave tables and passed through the dequantization unit, followed by the reduction unit and then the quantization unit \ding{190}. These stages are pipelined to hide latency. The scale factors generated for each wave by the quantization unit are transmitted immediately after the corresponding data, packaged into one or more dedicated scale packets. For example, with a 4 KB wave, a BlockSize of 64, and a maximum packet payload of 128 B, the scale factors for one wave occupy exactly one full packet.

\subsection{Scalability of SCIN}
\label{sec:Scalability}
\begin{figure}
    \centering
\includegraphics[width=\linewidth]{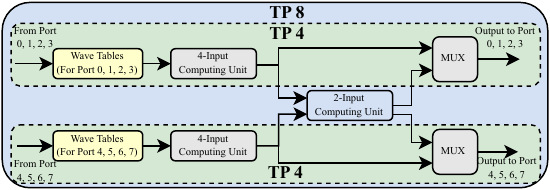}
\caption{Flexible support for one TP8 or two concurrent TP4 configurations within a centralized ISA.}    \label{fig:6}
\end{figure}
Currently, large language model (LLM) inference commonly employs 4-GPU or 8-GPU tensor parallelism \cite{agrawal2024vidur,agrawal2023sarathi,yang2025contextparallelism}. Meanwhile, the size of a single node has already reached 72 GPUs, as exemplified by NVIDIA Rubin NVL72 \cite{nvl72} and AMD Helios \cite{amd_helios}, and is expected to increase further in the future \cite{bhargava2026verarubinpod}. Therefore, this provides substantial headroom for scaling TP within the first-layer network, even as model sizes continue to grow. Meanwhile, because interconnect bandwidth typically decreases as the network hierarchy becomes deeper, model inference spanning multiple network layers generally favors pipeline parallelism or data parallelism, which are less sensitive to bandwidth \cite{agrawal2023sarathi,ubmesh,yang2025contextparallelism}.
Accordingly, we focus our scalability analysis of SCIN on a single-level network, consistent with UALink, whose in-network computing scheme is also defined within the first network layer.
Nevertheless, SCIN could potentially be extended to multi-layer networks by adopting an approach similar to a reduction tree \cite{7830486}.

Within this scope, SCIN provides both resource scalability and configuration flexibility despite its centralized ISA design, which uses a single compute unit rather than the per-port accelerators adopted by NVLS and UALink.
As illustrated in Fig.~\ref{fig:6}, a single ISA in an 8-GPU system supports either one TP8 configuration or two concurrent TP4 configurations.
Specifically, the resources provisioned for an 8-input All-Reduce can be partitioned into two sets, each supporting a 4-input All-Reduce operation.
This partitioning allows the wave table and compute unit, which dominate the hardware overhead, to be reused across different TP configurations, while enabling the two TP4 operations to execute concurrently without resource conflicts.
More generally, as the ISA scales to support larger systems, the resources provisioned for larger All-Reduce operations can also be reused to support multiple smaller operations.
SCIN can therefore flexibly support common TP configurations with low hardware overhead.
\subsection{FPGA-based Prototype of SCIN}
\label{sec:Prototype}
\begin{figure}
    \centering
\includegraphics[width=0.5\linewidth]{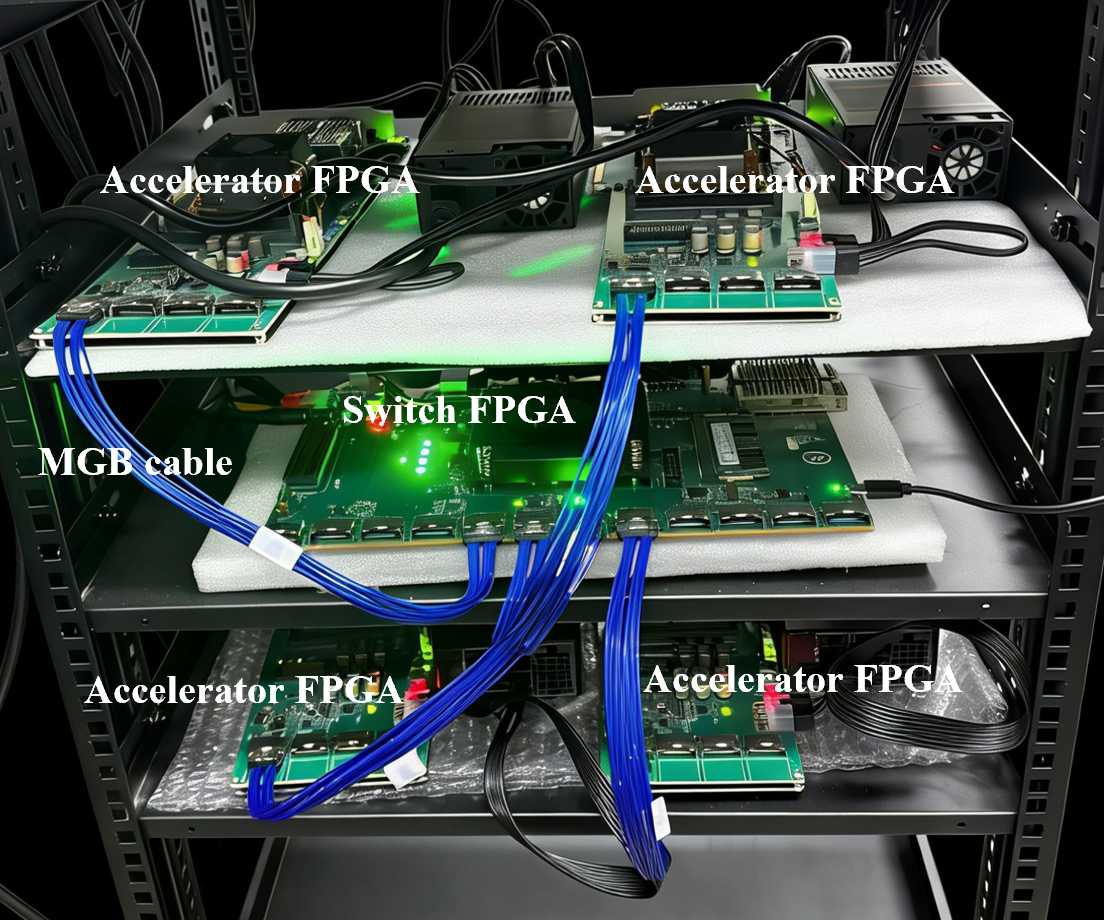}
    \caption{Photograph of the SCIN prototype.}
    \label{fig:7}
\end{figure}

The prototype implements the architecture shown in Fig.~\ref{fig:4} and consists of four endpoint FPGAs, which emulate accelerators, and one switch FPGA.
A photograph of the prototype is shown in Fig.~\ref{fig:7}. We use the AMD Aurora IP \cite{AURORA} to implement the physical and link layers, while the transport layer is organized as four types of buffers, each dedicated to a different class of memory transactions with credit-based flow control. On the accelerator side, the NoC adopts the AXI protocol \cite{AXI} and the INC flag is carried in the AXI user signals. 
We develop a protocol translation module between the NoC and the transport-layer buffers, similar to that in SUE \cite{SUE}. The module packs AXI requests and responses from the NoC into the corresponding transport-layer queues and, conversely, converts received memory transactions into AXI transactions for reading from or writing to accelerator memory (DDR4).
On the switch side, the ISA communicates directly with the transport-layer queues through handshake signals, avoiding unnecessary protocol-conversion latency. We use the MicroBlaze CPU to emulate GPU-side synchronization operations. 
The ISA instructions are preconfigured through a JTAG interface. 
In this system, the switch communicates with each endpoint through a single link, where each link consists of four Gigabit Transceiver (GT) lanes. Each lane provides 32 Gbps bidirectional bandwidth, resulting in an aggregate bidirectional bandwidth of 128 Gbps per link. The FPGA boards are connected using Multi-Gigabit Board (MGB) connectors \cite{MGB}. A 32B flit size is used, and accordingly the system operates at 250 MHz to sustain full bandwidth. The packet header occupies one flit, and the maximum packet payload is 4 KB, which maps exactly to one full AXI burst transfer.
The wave size is set to 4 KB.

In this prototype, the endpoint-to-switch latency is approximately 360 ns, consisting of 20 cycles of transport-layer latency and 70 cycles of latency in the link and physical layers introduced by the Aurora IP. We achieve an All-Reduce latency of 3.34 $\mu\text{s}$ for 4 KB messages and 2.27 ms for 16 MB messages.
For large messages, the achieved All-Reduce efficiency factor is 92.4\%. Profiling shows that the remaining unutilized bandwidth mainly comes from three sources:
1) the 64b/66b encoding and link-maintenance overhead in the Aurora IP ($\sim 3\%$),
2) non-ideal circuit effects such as pipeline bubbles between AXI transactions ($\sim 3\%$), and
3) protocol overhead ($\sim 1\%$). 
All devices are AMD XCVU13P-FLGA2577-2-I chips. The switch uses 8.8\% LUT, 4.8\% FF, 21.3\% BRAM, 0.59\% DSP, and 12\% GT, while the ISA alone uses 2.5\% LUT, 0.7\% FF, 1.2\% BRAM, and 0.59\% DSP.

\section{Evaluation}
\subsection{Methodology}
\label{sec:Methodology}

\noindent\textbf{Hardware-Calibrated Network Simulator.}
Although our FPGA prototype demonstrates the feasibility and efficiency of SCIN, its scale is limited. We therefore build a cycle-level network simulator based on BookSim2~\cite{jiang2013detailed} to evaluate SCIN in a larger, production-scale configuration. The simulated system comprises eight accelerators interconnected by four switches using the same topology as the NVIDIA DGX H200~\cite{H200}. We extend the BookSim2 router model to implement the ISA as described in Section~\ref{sec:isa}. Each per-accelerator wave table in the ISA is provisioned with 96~KB of storage with a wave size of 4 KB. Transport-layer queues for different memory-transaction types are mapped to separate virtual channels (VCs), which are sized to cover the round-trip latency between the switch and accelerators.
We parameterize the simulated NVLink and NVSwitch components using real device characteristics. The maximum packet payload is 128~B, and each packet header occupies 16~B, corresponding to one flit. Read requests and write responses are modeled as single-flit packets. 
The aggregate bidirectional bandwidth between each GPU and the switches is 900~GB/s, and the one-way GPU-to-switch latency is 250~ns.

For the NVLS baseline, following NVIDIA's NVLS design~\cite{klenk2020network}, we implement DMA-based NVLS by extending the BookSim2 router with NVLS-like in-switch reduction and multicast operations. Each switch port is provisioned with a 320~KB reduction table organized into 1~KB entries, as discussed in Section~\ref{sec:isa}. We also configure the baseline with a $2\times$ switch internal speedup, as adopted by the NVLS design, to mitigate the bandwidth pressure caused by in-network multicast. SCIN requires no such internal speedup.

\noindent\textbf{Validation.}
We validate simulator fidelity by comparing simulated NVLS All-Reduce performance with NCCL measurements \cite{bachan2025nccl227} on real hardware for message sizes from 32~KB to 1~GB. The simulated results closely match the hardware measurements, with an average error of 5.5\% and an error of 2.6\% for large messages.
We also calibrate the simulated SCIN against our FPGA prototype, with discrepancies below 1\%. 

\noindent\textbf{Profiling-Based Compute Simulator.}
In TP inference, computation and communication typically do not overlap and can therefore be simulated independently. To evaluate the impact of SCIN on end-to-end LLM inference latency, we develop a TensorRT-LLM-based profiling tool~\cite{tensorrtllm} to obtain realistic GPU execution latencies. We profile LLaMA-2 models (7B, 13B, and 70B)~\cite{llama2} and Mixtral-8$\times$7B~\cite{jiang2024mixtralexperts} on an H200 GPU using a production-like execution path, with both input and output sequence lengths set to 1024 tokens.
We use batch sizes below 128, representing low-concurrency scenarios in which TP typically achieves lower latency and higher throughput than other parallelization strategies \cite{wang2026moebiusservingmixtureofexpertmodels,tan2025vllmmoe}.
To emulate TP execution, we shard the attention heads and MLP intermediate dimensions according to the TP size, ensuring that the profiled per-GPU compute latency reflects practical TP deployments. We then combine these compute profiles with the simulated communication latency to estimate end-to-end inference performance.

\noindent\textbf{Quantization Evaluation Setup.}
For quantization evaluation, we implement our evaluation framework on top of the open-source SmoothQuant codebase \cite{xiao2023smoothquant}. We simulate the computation path of quantized All-Reduce in software to evaluate its impact on model accuracy. The data involved in All-Reduce is symmetrically quantized. We evaluate the impact of applying INQ All-Reduce to both FP16 models and FP8-quantized models and also compare the accuracy with ring-based quantized (RQ) All-Reduce. For FP8 quantization, both weights and activations are quantized symmetrically on a per-tensor basis, while activations are quantized dynamically.
INQ All-Reduce is evaluated on several LLM models with different architectures, including LLaMA-2 \cite{llama2}, Mistral \cite{jiang2023mistral7b}, and Mixtral \cite{jiang2024mixtralexperts}, where Mixtral is a Mixture-of-Experts (MoE) model. Model accuracy is measured on multiple datasets, including the Massive Multitask Language Understanding (MMLU) \cite{mmlu} in the five-shot setting and zero-shot Commonsense QA benchmarks such as WinoGrande \cite{WINO}, PIQA \cite{PIQA}, HellaSwag \cite{HS}, ARC \cite{arc1}, BoolQ \cite{clark2019boolqexploringsurprisingdifficulty} and OBQA \cite{OPEN}. All evaluations are conducted using the Language Model Evaluation Harness \cite{eval}.
\subsection{In-Network Quantized All-Reduce}

\label{sec:INQ}

\begin{table}
\centering

\caption{Comparison of perplexity for the LLaMA-2-7B model under FP16 precision across different block sizes when applying RQ All-Reduce or INQ All-Reduce separately (TP=8).
Results are evaluated on the WikiText-2 dataset \cite{wiki2} with a sequence length of 2048. Lower perplexity is better.}
\label{table:1}
\begin{tabular}{ccccccl}
\hline
\multirow{2}{*}{\textbf{Method}} & \multirow{2}{*}{\textbf{Bits}} & \multicolumn{5}{c}{\textbf{Block Size}}    \\ \cline{3-7}
                                 &                                & 32     & 64     & 128    & 256    & 512    \\ \hline
FP16                             &                                & \multicolumn{5}{c}{5.4721}                 \\ \hdashline
RQ                           & INT8                           & 5.4742 & 5.4755 & 5.4757 & 5.4839 & 5.5050 \\
INQ                           & INT8                           & 5.4736 & 5.4746 & 5.4741 & 5.4815 & 5.5020 \\\hdashline
RQ                            & INT4                           & 5.7977 & 5.9782 & 6.5894 & 7.7286 & 14.1795 \\
INQ                          & INT4                           & 5.6055 & 5.6695 & 5.8628 & 6.1649 & 7.1776 \\ \hline
\end{tabular}
\end{table}

\begin{table*}
\centering
\caption{Accuracy (\%) of 8-bit INQ All-Reduce on FP16 and FP8 Models ($\text{TP}=8$, $\text{Block Size}=64$)}
\label{table:2}
\resizebox{0.84\textwidth}{!}{%
\begin{tabular}{cccccccccccccc}
\hline
\multirow{2}{*}{\textbf{Method}} & \multicolumn{5}{c}{\textbf{MMLU}}                                                  & \multicolumn{8}{c}{\textbf{Commonsense QA}}                                                                                                 \\ \cline{2-14}
                                 & \textbf{Hums.} & \textbf{STEM} & \textbf{Social} & \textbf{Other} & \textbf{Avg.}  & \textbf{HellaSwag} & \textbf{PIQA} & \textbf{WinoGrande} & \textbf{ARC-e} & \textbf{ARC-c} & \textbf{BoolQ} & \textbf{OBQA} & \textbf{Avg.}  \\ \hline
LLaMA-2-7B                       & 43.25          & 37.11         & 51.84           & 52.75          & 45.86          & 76.01              & 79.11         & 69.14               & 74.58          & 46.25          & 77.71          & 44.20         & 66.71          \\
\textbf{FP16 + INQ AR}           & 43.17          & 37.23         & 51.45           & 52.94          & \textbf{45.81} & 76.06              & 79.05         & 69.14               & 74.37          & 46.42          & 77.83          & 44.00         & \textbf{66.70} \\ \hdashline
FP8                              & 42.91          & 37.81         & 51.15           & 51.79          & 45.53          & 75.71              & 78.89         & 69.14               & 73.91          & 45.14          & 77.65          & 44.20         & 66.38          \\
\textbf{FP8 + INQ AR}            & 42.38          & 37.65         & 50.76           & 52.53          & \textbf{45.40} & 75.85              & 78.78         & 68.82               & 74.54          & 45.56          & 77.74          & 43.80         & \textbf{66.44} \\ \hline
LLaMA-2-13B                      & 53.45          & 43.77         & 63.18           & 61.47          & 55.18          & 79.38              & 80.52         & 72.14               & 77.40          & 49.06          & 80.55          & 45.20         & 69.18          \\
\textbf{FP16 + INQ AR}           & 53.05          & 44.15         & 63.15           & 61.80          & \textbf{55.20} & 79.37              & 80.52         & 71.98               & 77.61          & 49.23          & 80.55          & 45.60         & \textbf{69.27} \\ \hdashline
FP8                              & 51.48          & 43.61         & 62.79           & 61.60          & 54.43          & 79.15              & 79.82         & 71.90               & 77.40          & 49.15          & 80.24          & 46.60         & 69.18          \\
\textbf{FP8 + INQ AR}            & 52.14          & 43.55         & 62.98           & 61.67          & \textbf{54.69} & 79.28              & 80.09         & 71.67               & 76.77          & 48.98          & 80.00          & 46.20         & \textbf{69.00} \\ \hline
Mistral-7B                       & 56.83          & 52.49         & 73.35           & 70.49          & 62.50          & 81.09              & 82.21         & 74.11               & 79.46          & 53.84          & 83.76          & 44.20         & 71.24          \\
\textbf{FP16 + INQ AR}           & 56.64          & 52.93         & 72.83           & 70.10          & \textbf{62.33} & 80.93              & 82.21         & 74.74               & 79.46          & 54.01          & 83.52          & 44.80         & \textbf{71.38} \\ \hdashline
FP8                              & 55.64          & 52.81         & 72.41           & 69.17          & 61.67          & 80.80              & 81.94         & 74.27               & 79.04          & 53.33          & 83.00          & 43.40         & 70.82          \\
\textbf{FP8 + INQ AR}            & 55.88          & 52.55         & 72.18           & 69.55          & \textbf{61.73} & 80.81              & 81.94         & 73.95               & 79.17          & 53.67          & 83.39          & 44.00         & \textbf{70.99} \\ \hline
Mixtral-8x7B                     & 64.80          & 61.47         & 80.86           & 77.66          & 70.42          & 84.00              & 83.62         & 76.48               & 83.59          & 59.56          & 85.29          & 46.20         & 74.10          \\
\textbf{FP16 + INQ AR}           & 65.06          & 61.47         & 81.05           & 77.44          & \textbf{70.50} & 84.00              & 83.62         & 75.69               & 83.75          & 59.81          & 85.26          & 47.40         & \textbf{74.22} \\ \hdashline
FP8                              & 64.36          & 61.47         & 81.25           & 76.25          & 70.04          & 83.98              & 83.35         & 75.61               & 82.91          & 60.15          & 85.41          & 48.20         & 74.23          \\
\textbf{FP8 + INQ AR}            & 64.21          & 60.70         & 81.05           & 76.50          & \textbf{69.83} & 83.90              & 83.35         & 75.14               & 83.21          & 59.22          & 85.38          & 46.40         & \textbf{73.80} \\ \hline
\end{tabular}
}
\end{table*}

Prior work \cite{lee2024exploring,xiao2023smoothquant} shows that 8-bit activation quantization can often preserve model accuracy. However, most studies focus on quantizing complete activations, while the accuracy impact of independently quantizing partial activations before they are combined by All-Reduce remains less well understood. We investigate this setting in our work.
We first compare INQ All-Reduce with RQ All-Reduce. Table \ref{table:1} reports the perplexity (PPL) results under different quantization bit widths and block sizes with TP=8. When All-Reduce is quantized to INT8, INQ All-Reduce preserves nearly the same perplexity as the FP16 baseline, with only a 0.0025 PPL degradation when the block size is 64, and consistently outperforms RQ All-Reduce. At INT4, INQ All-Reduce still incurs acceptable perplexity degradation at smaller block sizes and significantly outperforms RQ All-Reduce.
These results indicate that our quantization method is more robust than RQ All-Reduce under more challenging quantization settings, and therefore is likely to generalize better on more difficult quantization tasks. 
Moreover, because SCIN requires only a single quantization step regardless of TP size during communication, it is expected to scale better than the ring-based method at larger TP sizes.

In this work, we choose to quantize All-Reduce to INT8 with a block size of 64 in order to achieve near-lossless compression and a 1.94x communication-compression ratio. Table \ref{table:2} presents the evaluation results of 8-bit INQ All-Reduce on several LLMs with a block size of 64 and TP=8.
These results demonstrate that, across all evaluated datasets, INQ All-Reduce generalizes well across models with diverse architectures and numerical precisions, including both full-precision and FP8-quantized models, while introducing negligible accuracy degradation.
In some cases, slight improvements are observed---consistent with expectations when applying activation quantization in LLMs \cite{lee2024exploring}.

\subsection{All-Reduce Bandwidth and Latency Improvement}
\begin{figure}
    \centering

    \begin{subfigure}{\linewidth}
        \centering
        \includegraphics[width=0.75\linewidth]{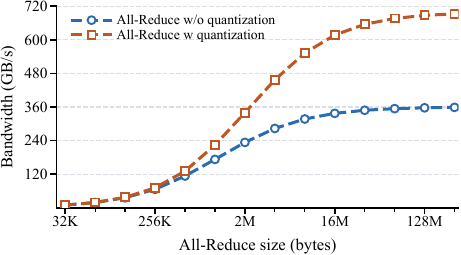}
        \caption{SCIN All-Reduce bandwidth}
        \label{fig:8_a}
    \end{subfigure}

    \begin{subfigure}{\linewidth}
        \centering
        \includegraphics[width=0.8\linewidth]{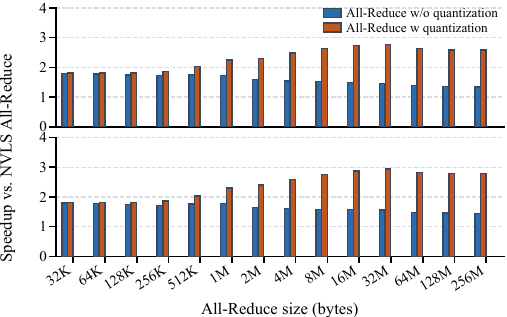}
        \caption{Speedup over NVLS for 8-GPU (top) and 4-GPU (bottom).}
        \label{fig:8_b}
    \end{subfigure}

    \caption{Simulated All-Reduce performance with 24 waves per table, each wave with 4 KB of data and 128 B of scale.}
    \label{fig:8}
\end{figure}

The All-Reduce bandwidth is defined as the message size divided by the simulated execution time, independent of the number of GPUs. The maximum achievable unidirectional payload bandwidth is 360 GB/s due to the packet header overhead. Fig.~\ref{fig:8_a} reports the All-Reduce performance of SCIN, including barrier overhead. For quantized All-Reduce, we define its bandwidth as the equivalent bandwidth that an unquantized All-Reduce would require to achieve the same latency. For large messages, SCIN approaches the maximum achievable payload bandwidth with 96~KB of data capacity and an additional 3~KB of scale-factor storage per wave table. This performance stems from SCIN's in-network multicast, which avoids crossbar contention, together with its wave-regulation mechanism, which hides synchronization overhead while efficiently utilizing buffer resources.

On the eight-GPU system, SCIN achieves up to a $1.8\times$ speedup over NVLS for small messages by eliminating redundant data traversal and reducing synchronization latency. For large messages, unquantized SCIN achieves a $1.34\times$ speedup over NVLS. Enabling quantization further reduces the communication volume, increasing the equivalent bandwidth by $1.94\times$ over unquantized SCIN and yielding up to a $2.58\times$ speedup over NVLS. The highest speedup, $2.76\times$, occurs at medium message sizes of 16~MB and 32~MB, where SCIN benefits from both lower small-message latency and the bandwidth savings offered by quantization. We also evaluate SCIN on a four-GPU system, where its performance remains nearly identical to that on the eight-GPU system because SCIN does not incur scale-dependent bandwidth-efficiency loss. In contrast, according to Equation~\eqref{eq:nvls_bandwidth_efficiency}, the bandwidth-efficiency factor of NVLS decreases from 0.89 with eight nodes to 0.80 with four nodes. Consequently, SCIN's large-message speedup over NVLS increases to $2.78\times$.

\subsection{Hardware Overhead Analysis}
\label{sec:Hardware}
\begin{figure}
    \centering
\includegraphics[width=0.65\linewidth]{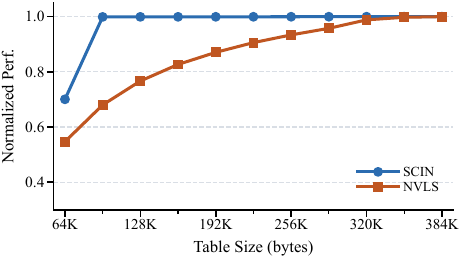}
    \caption{Performance sensitivity to reduction table size.}
    \label{fig:9}
\end{figure}
\label{sec:Regulation}

In-network multicast can impose substantial load on the switch crossbar. In SCIN, multicast packets are routed directly to the output ports through dedicated hardwired datapaths. These datapaths isolate different multicast streams, eliminating crossbar contention among them and avoiding the resulting network congestion. However, in NVLS, in-network multicast packets generated at different ports must traverse the shared crossbar before reaching the destination GPUs. Following NVIDIA's NVLS design, we provision the crossbar with a $2\times$ internal speedup. Nevertheless, multicast-induced congestion can still occur, as reported in the original NVLS study~\cite{klenk2020network}. Consequently, NVLS performance is more sensitive to reduction-buffer capacity. As shown in Fig.~\ref{fig:9}, SCIN approaches peak bandwidth with a relatively small 96~KB wave table, which functions as the reduction buffer. In contrast, NVLS requires approximately twice its theoretical minimum buffer capacity to achieve peak bandwidth. This observation is consistent with the experimental results reported in the original NVLS study~\cite{klenk2020network}.

In addition to the FPGA resource utilization reported in Section~III, we synthesize our design using the TSMC 28-nm process to evaluate its ASIC area overhead. The switch-side modifications occupy approximately 4.04~mm\textsuperscript{2}. Even without accounting for technology scaling, this area is less than 2\% of the 297~mm\textsuperscript{2} die area of NVIDIA's NVLink4 NVSwitch  \cite{ishii2022nvlink}, which is fabricated using a 4-nm process. SCIN introduces no appreciable hardware area overhead on the GPU side. These results confirm that our proposed architectural support is cost-effective and hardware-feasible.

\subsection{End-to-End Inference Speedup}
\begin{figure}
    \centering
\includegraphics[width=0.9\linewidth]{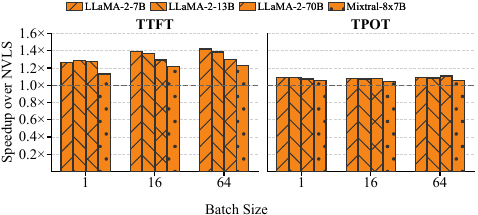}
\caption{End-to-end TTFT and TPOT speedups of SCIN over NVLS for LLaMA-2 and Mixtral models at FP8 inference precision with $\mathrm{TP}=8$.}  
\label{fig:10}
\end{figure}

Fig.~\ref{fig:10} reports the speedup of SCIN over NVLS for TP inference on LLaMA-2 and Mixtral-8$\times$7B. SCIN improves TTFT and TPOT by up to $1.42\times$ and $1.12\times$, respectively.
SCIN achieves a larger TTFT speedup because prefill is generally more communication-intensive than decode. The TTFT speedup also increases with batch size: communication volume grows linearly with batch size, whereas computation benefits from improved GPU utilization at larger batches. Consequently, communication time grows faster relative to computation time, increasing its contribution to end-to-end latency. During decode, communication accounts for at most 19\% of execution time in our evaluated workloads. Amdahl's Law therefore limits the maximum TPOT speedup to $1.12\times$. 
We expect this benefit to increase as compute hardware advances faster than interconnect bandwidth and software optimizations further reduce computation time. For example, SiFAR~\cite{taneja2026sifarsynchronizationfreeallreducelowlatency} reports that NVLS All-Reduce accounts for 30\% of decode time after reducing computation overhead through megakernel optimizations, suggesting greater potential benefit from SCIN.
Moreover, latency-sensitive agentic workloads may favor larger TP degrees, exposing more communication overhead and increasing SCIN's benefit.
Mixtral-8$\times$7B benefits less than comparable dense models because expert outputs are locally aggregated before inter-GPU All-Reduce, reducing TP communication's share of end-to-end latency.

\section{Related Work}
To mitigate the bottleneck of collective communication, many prior works have introduced specialized hardware into switches to accelerate collective operations \cite{klenk2020network,7830486,UALink,li2019accelerating,zhang2026towards,sapio2021scaling,101145,lao2021atp,MLSYS2021_5c6614ea,9910089}.
However, most works focus on traditional inter-node networks.
Only NVLS \cite{klenk2020network}, DySHARP~\cite{zhang2026acceleratingmoedynamicinswitch}, UALink \cite{UALink}, CAIS \cite{zhang2026towards}, and TRACI \cite{101145} target inter-GPU interconnects with memory semantics. 
In addition to NVLS's pull-based reduction, the original NVLS study~\cite{klenk2020network} also proposed a push-based architecture, which CAIS builds upon. However, this architecture requires an eviction mechanism that evicts in-progress reductions from the switch to a GPU to avoid deadlock, increasing implementation complexity. Moreover, it also retains an address-based, element-granular reduction abstraction and therefore faces the same fundamental limitation as NVLS in supporting scale-aware quantized in-network reduction.
DySHARP extends NVLS with dynamic in-network broadcast and reduction to accelerate All-to-All communication in expert parallelism. 
As an NVLS-based design, it inherits the architectural limitations identified in this work.
TRACI is specifically designed for DLRM workloads.
More recently, UALink 2.0 defines support for in-network computing (INC). However, this standard does not provide a detailed architectural implementation for its in-network computing and also does not define support for scale-aware quantized All-Reduce. Moreover, its per-port INC-engine organization potentially causes multicast-induced network congestion.
Many approaches mitigate All-Reduce overhead in LLM inference through software optimizations \cite{gond2026tokenweaveefficientcomputecommunicationoverlap,taneja2026sifarsynchronizationfreeallreducelowlatency,zhu2025nanoflowoptimallargelanguage}. These techniques are often orthogonal to SCIN, which optimizes the communication operation itself.
\section{Conclusion}
We propose SCIN, a switch-centric in-network architecture for shared-memory AI-accelerator fabrics. SCIN eliminates NVLS's redundant data movement and synchronization overhead while enabling quantized All-Reduce that nearly halves communication volume with negligible accuracy loss. Our evaluation demonstrates substantial All-Reduce and end-to-end inference speedups over NVLS.

\bibliographystyle{ACM-Reference-Format}
\bibliography{sample-base}

\end{document}